\documentclass[12pt]{article}
\usepackage{graphicx}
\begin{document}
\centerline{\large \bf Introduction to Statistical Physics outside Physics}

\bigskip

\centerline{Dietrich Stauffer}

\centerline{Institute for Theoretical Physics, Cologne University}

\centerline{D-50923 K\"oln, Euroland}

\bigskip
\noindent
Abstract:  We review the possibilities and difficulties for statistical
physicists if they apply their methods to biology, economics, or sociology.
WARNING: I report opinions, not simulations.


\medskip

\section{Introduction}

The basic theorem of interdisciplinary research states: Physicists not only 
know everything; they know everything better.
This theorem is wrong; it is valid only for 
computational statistical physicists like me. (Some conference speakers
still seemed to believe in the theorem; they can discuss it in their papers.)
Here I present my personal view about the application of statistical physics
(simulation) methods to fields outside physics like biology, economic, 
sociology and even psychology.

\section{ Possible criticism}

Typical criticisms levelled against physicists outside physics are
"Biology is more complicated than physics, physicists neglect details." True;
but the Earth is more complicated than a point mass; nevertheless Kepler's laws
are quite accurate and were very useful. Physicists should try to find the 
simplest model giving the desired result, and not to make the model more 
realistic only for the sake of realism. Of course, Kepler knew that the Earth 
was not a point. He knew of the 30-year war destroying central Europe during 
his later life, for he was an unlucky astronomer advising general Wallenstein
who was murdered. Kepler died when he tried to get his astrologer's salary from
the imperial parliament. So, Kepler could have tried to build the large size
of the Earth into his description, to deal with the different countries of
Europe, and to understand the religious differences which caused the war to 
start. But this would have been a hindrance, not a help, for the understanding 
of planetary motion. Only much later, thanks to Newton and others, did mankind
understand that a spherically extended mass has a gravitational field
equal to that caused if the whole mass would have compressed to a point in the
center of the sphere. And even later, thanks to Einstein's General Relativity 
Theory, did experts understand how the mass affects the metric of time and 
length through curved space. For the planetary motion, Einstein's theory then 
explained minor deviations from Kepler's laws like the rotation of Mercury's 
perihel. Perhaps the now started century will explain how the Higgs particle
produces the mass, and bring new effects. But the basic Kepler laws from 
around 1610 are still the starting point of modern physics. Of course, if you
want to be a geographer, the assumption of a point-like Earth is dangerous 
for your employment: The same model may be good for some and bad for other
purposes.
 
"Computers know only 0 and 1." Wrong; many shades of grey are possible, and
integers on 32-bit computers vary between -2147483648 and +2147483647. I don't
know a human being who can distinguish anything with an accuracy of $10^{-9}$.
But when dealing with a new problem, why not first assume that the state
variables can only be $+1$ and $-1$ (1 and 0 for informatics). Monte Carlo
simulations with discrete Ising models can be done for thousand times bigger 
systems than in molecular dynamics studies on a continuum ($10^{13}$ versus
$10^{10}$). Whoever studied
newspapers in the last decades came across the difficulties of defining when
a person is dead, allowing the organs to be transplanted. Our hair and nails 
still grow long after our hearts stop beating; some readers of my papers 
claim my beer belly grew after my brain stopped working. Nevertheless, 
newspapers report the death of a person, and statistical offices prepare life
tables as if life ($+1$) and death ($-1$) are the only possibilities without
anything in between. It is animals like {\it Caenorhabditis elegans} with
its "dauer" state (some sort of hibernation) for which intermediate states 
are very important, not human beings. Finally, many social problems are 
connected with more than one variable, but so are many models of physicists.
Seven-dimensional percolation and five-dimensional Ising models show that
physics is not one-dimensional, and neither is immunology where a a shape
space of high dimensions was suggested already in 1979 before such models 
could easily be simulated. (In contrast to widespread belief, the Ising 
model does not only allow $+1$ and $-1$ as state variables. Much work has 
been done on Blume-Capel-Rys-Blume-Emery-Griffiths models where also zero
is allowed.)
 
"Humans are not numbers." Wrong; we just don't {\it want} to be treated as 
numbers,
and prefer a name tag instead of a number for identification. But reality is
different: US citizens have a social security number, I have a national 
identity card with a computer-readable number; only with a 
computer-readable passport can I now enter the USA; my employer, my health 
insurance and my bank all gave me a number. Whether I smoke, drink w\'odka in 
the morning, and eat steaks every evening influences my date of death, and 
neither employer nor health insurance know about it. Nevertheless, by averaging
over millions of people, these personal details cancel out. Similarly, the
decision to get a child is highly private, but nevertheless I am most likely
right if I claim that in 2004 hundreds of thousands of children will be born
in Germany. If today's pension
plans are in trouble it is because the {\it average} life expectancy increases 
and the {\it average} number of births per women decreased, not because of 
individual fluctuations. Thus humans and atoms may be described by the same 
method, if we look at averages. And besides the above rather recent examples
we have astronomer Halley trying to establish the first human life tables
three centuries ago, and Greek philosopher Empedokles claiming more than two
millennia ago (according to Mimkes) that some people are like water and wine, 
mixing easily, while others are like water and oil, not mixing. 
 
"Physics is different from Sociology and/or Economy." 
True; but Schelling [1] in first issue 
of J. Mathematical Sociology simulated something like the dilute Ising model 
with Kawasaki kinetics to explain black ghettos in USA; and Nobel laureate 
Stigler [2] published market simulations in 1964 before any conference 
participant wrote the first simulation paper. So why cannot we follow them?
Indeed, our problem is much more than we may not know such pioneering papers
in non-physics journals, and thus erroneously believe that we bring new concepts
to such fields with our physics experience.

\section{Do other fields welcome physicists?}

From my own experience in bio-, econo- and socio-physics my answer depends on
the field; psychophysics [3] is too new for me.
 
\subsection{Economics: yes}

Economy Nobel laureate Harry Markowitz [4] wrote:  ``I believe that 
microscopic market simulations have an important role
to play in economics and finance. If it takes people from outside economics
and finance -- perhaps physicists -- to demonstrate this role, it won't be
for the first time that outsiders have made substantial contributions to these
fields.'' Economics professor Thomas Lux knows a lot about physics; one we 
discussed whether to do a Grassberger-Procaccia analysis of some financial data
and then I realized that in this case he, the economist, would have to explain 
to me, the physicist, how to apply this physics method. Economics professor Haim
Levy wrote papers and a book with theoretical physicist Sorin Solomon and with
his son Moshe Levy, a doctoral student of physics who is now economics 
professor.

Kert\'esz reported that at a recent econophysics conference the economists 
criticized the physicists for re-inventing the economic wheels. This I count as
acceptance, not as rejection: If A claims do have done first what B reports 
now, then A regards this B problem as important. Somewhat similarly, when in our
university magazine I reported about Ehrenstein's simulations of the Tobin tax,
a senior economics professor criticized me, perhaps correctly, for having 
ignored the dangers of government currency controls, as Germany had it during 
the Nazi dictatorship; again, the problem was taken serious by an economist.
  
\subsection{Sociology: yes, if computational sociologists}

A year after I started to simulate Sznajd models, I presented a review at the
fifth conference on simulating society, SIMSOC V, in Poland, organized by
psychologist A. Novak. I felt quite at 
home with some of the talks of non-physicists, with simple models giving
clear results. In contrast the international Physics Computing 2001 meeting
two weeks earlier was full of complicated simulations understandable to 
specialists only. I learned from Hegselmann's talk (from a philosophy 
department) about bounded confidence and used that in several later papers of 
mine. Of course, for a conference on "Simulating Society" the participating 
sociologists are a highly biased selection.

An important problem is how to get data, which then should be modeled by
sociophysics. Election results, more precisely the distribution of votes among
many candidates, have been modeled successfully for both Brazil and India. This
example [5] also shows the limitations: The methods of statistical physics can
predict the probability distribution functions for the velocity of air molecules
and the election votes, but they cannot predict where a specific air molecule 
will be a second later or which candidate will win the election. The internet
allows physicists and computer scientists to collect lots of data on who is
sending e-mail to whom, which web site cites another web site, and which 
computers are directly connected to other computers. These evolving networks 
yield much better statistics than the less than 100 people at the South Pole
recently investigated for social roles and leadership [6], ignoring the recent
network literature in physics.

\subsection{Biology: no, but changing} 

Traditionally, the only good biology was experimental biology, a view also held
by famous interdisciplinary physicist H.E. Stanley (whom I had to replace by
this talk) who analyzed empirical data
from biology and economics. But with billions of base pair data available
now from DNA analysis of the whole genome, the need to interpret these 
biological data arises. Spin glass physics taught us that knowing all the 
interactions between the spins and all the spin orientations does not yet solve
the problem how the spin glass behaves as a whole. Similarly, bio-informatics 
needs to understand how to deduct from the known DNA the behaviour of the 
organism. Today, nobody has as yet produced a living being out of non-living
matter. Thus biology should take modelling more serious than in past decades,
and I see some changes in the behaviour of young biologist. After all, I 
became a member of the advisory board of one biological and one medical 
journal, never having followed a university course in these fields.

What we need in biologically motivated simulations are real successes. There are
lots of papers on theoretical immunology, but vaccination works since more than
a century without needing my excellent simulations. Ageing still cannot be
prevented, in spite of experimental and theoretical work on it. A big effect, 
comparable to the atomic bomb of 1945, is still lacking in theoretical biology.
 
\subsection{Econo-bio-socio-physics ?}

Of course, even more difficult is to become accepted by work touching on all
three of the above fields, like predicting the "age quake" around 2030 when
the 70-year old people may be the strongest age cohort in rich countries. 
Should they trust existing pension regulations or build up their own savings
for old age? Politicians debate this hotly in several countries of the 
European Union. Thus far none of them asked for my advice.

\section{Acceptance time}   

Let us {\it assume} that these interdisciplinary attempts of physicists are
objectively good. How long do we have to wait until they will be accepted
by the other field? After one decade of ageing simulations by physicists, a
few biologists have started to cite them. A dozen years after the car traffic
models of Nagel and Schreckenberg as well as Biham, Middleton and Levine, I
don't see yet applications in many cities and on many expressways, but serious
application effort exists in Duisburg (Germany), Dallas (Texas) and Portland 
(Oregon). In 1976 the later Nobel laureate de Gennes was one of those who 
suggested that the critical exponents of three-dimensional percolation should 
be found in real gelation experiments. This suggestion was far from 
revolutionary since the later chemistry Nobel laureate Flory invented 
percolation theory in 1941 to explain gelation (Bethe lattice, same universality
class as random graphs or other mean-field approaches). Nevertheless it took
more than a decade before chemistry experimentalists widely accepted this
idea. Thus, for fields much further apart, like physics and sociology, we should
measure the times for possible acceptance in decades, not in months. Similarly,
people planting trees for a forest work for future generations.

\section{Summary}
Exotic physics has become less exotic at this conference but it still rejected
by many, both within and outside physics departments. But there is some 
consolation:
The gap between theoretical physicists and theoretical biologists (sociologists)
may be smaller than that between the latter ones and experimental biologists
(sociologists). 

\bigskip
\newpage
\parindent 0pt
[1] T.C. Schelling, J. Mathematical Sociology 1 (1971) 143.

[2] G.J. Stigler, Journal of Business 37 (1964) 117.

[3] L. da Fontoura Costa, cond-mat/0309266.
 
[4] S. Moss de Oliveira, P.M.C. de Oliveira, and D. Stauffer, {\it Evolution,
Money, War and Computers} (Teubner, Stuttgart-Leipzig, 1999)

[5] M.C. Gonzalez, A.O. Sousa, and H.J. Herrmann, Int. J. Mod. Phys. C 15,
(2004) issue 1.

[6] J.C. Johnson, J.S. Boster and L.A. Palinhas, J. Mathematical Sociology 27
(2003) 89
\end{document}